\documentclass[a4paper,11pt,twocolumn]{article}
\usepackage{cite}
\usepackage[T1]{fontenc}
\usepackage{graphicx}
\usepackage{authblk}
\usepackage{amsmath}
\usepackage[margin=1in]{geometry}

\title{Low-noise octave-spanning mid-infrared supercontinuum generation in a multimode chalcogenide fiber}
\date{}

\author[1]{\small Zahra Eslami}
\author[1]{\small Piotr Ryczkowski}
\author[1]{\small Lauri Salmela}
\author[1]{\small Go\"ery Genty}
\affil[1]{\footnotesize Photonics Laboratory, Physics Unit, Tampere University, 33014 Tampere, Finland}
\begin{document}

\twocolumn[
\begin{@twocolumnfalse}
\maketitle
\begin{abstract}
We demonstrate the generation of a low-noise, octave-spanning mid-infrared supercontinuum from 1700 to 4800 nm by injecting femtosecond pulses into the normal dispersion regime of a multimode step-index chalcogenide fiber with 100 $\mu$m core diameter. We conduct a systematic study of the intensity noise across the supercontinuum spectrum and show that the initial fluctuations of the pump laser are at most amplified by a factor of three. We also perform a comparison with the noise characteristics of an octave-spanning supercontinuum generated in the anomalous dispersion regime of a multimode fluoride fiber with similar core size and show that the all-normal dispersion supercontinuum in the multimode chalcogenide fiber has superior noise characteristics. Our results open up novel perspective for many practical applications such as long-distance remote sensing where high power and low noise are paramount. 
\end{abstract}
\end{@twocolumnfalse}
]

Supercontinuum (SC) generation in the mid-infrared has attracted significant interest over the past decade due to its wide range of potential applications from tissue imaging \cite{petersen2018mid,seddon2013potential,dupont2012ir} and spectroscopy \cite{islam2009mid,eggleton2011chalcogenide} to defense and security \cite{kumar2012stand}. Supercontinuum sources are generally optimized in terms of spectral coverage or power for a particular application, but their noise characteristics are also of particular significance as large pulse-to-pulse fluctuations can severely degrade the performance and reduce the achievable resolution and contrast e.g. in sensing or imaging \cite{dupont2014ultra,gonzalo2018ultra,jensen2019noise}. Pulse-to-pulse instabilities in SC generation arise from nonlinear amplification dynamics of the input pulse fluctuations. These dynamics depend on various parameters including the pump pulse wavelength, energy, and duration \cite{newbury2003noise,gaeta2002nonlinear,moller2012power}, such that by carefully selecting the pump pulse characteristics one can in principle reduce the SC fluctuations \cite{dudley2002coherence}. More specifically, when pumping in the anomalous dispersion regime, intensity (and phase) fluctuations can be minimised using short pump pulses with femtosecond duration to seed the initial spectral broadening dynamics from the pump spectral components. In the case of longer pump pulses (typically few hundred femtosecond and beyond), the broadening mechanism is triggered by the noise present on top of the input pulse, which leads to significant shot-to fluctuations and generally high intensity noise.

The generation of a broadband supercontinuum in the mid-infrared requires using fibers made of soft glass with attenuation much lower than that of silica-based fibers, and in recent years there have been many studies of SC generation in fluoride \cite{li2019step}, chalcogenide \cite{diouf2017super,wang20171,diouf2019numerical,liu2016coherent}, multi-compound glasses \cite{jiao2019mid,nguyen2018highly,zhang2019ultrabroadband} and other type of soft glass fibers \cite{klimczak2014coherent}. Most of these studies, however, use single-mode photonics crystal fibers or tapered fibers with small core size to enhance the nonlinearity and engineer the dispersion properties, thereby limiting the injected power due to the low damage threshold of soft glasses. For mid-infrared applications where high power is especially crucial due to the lack of sensitive detectors, the possibility to generate a broadband SC in multimode fibers with larger core enabling to inject more power has recently attracted attention \cite{eslami2019high,swiderski2014high,zhang2016broadband,shi2016multi,khalifa2016mid}. These studies typically use anomalous dispersion pumping as this particular regime has been shown to yield the broadest SC spectra \cite{dudley2006supercontinuum,genty2007fiber}. However, this regime is also particularly sensitive to noise amplification dynamics, leading to large shot-to-shot intensity fluctuations \cite{dudley2006supercontinuum,genty2007fiber,klimczak2016direct} and the development of high-power broadband mid-infrared SC sources with low noise is thus still challenging. 

Here, we report the generation of a low noise, octave-spanning, mid-infrared supercontinuum in a multimode chalcogenide fiber using an all-normal dispersion pumping scheme. The all-normal pumping approach has the advantage to be less sensitive to noise amplification dynamics than the anomalous dispersion regime as the spectral broadening mechanism in this case arises mainly from self-phase modulation (SPM) allowing to preserve high stability even with few hundred femtosecond pump pulses \cite{dudley2006supercontinuum,genty2007fiber,hooper2011coherent}. We perform a systematic study of the noise performance by measuring the pulse-to-pulse intensity fluctuations in different wavelength bands across the SC spectrum and we compare the results with the noise characteristics measured for an octave-spanning supercontinuum generated in the anomalous dispersion regime of a multimode fluoride fiber with similar core size and length. Our study shows that pumping in the normal dispersion regime leads to the generation of a SC with significantly reduced noise level and opens novel perspectives for the generation of high-power broadband mid-infrared SC sources for applications such as remote sensing where the stability and power are more important than the beam profile.

\begin{figure}[t]
\centering
\includegraphics[width=\linewidth]{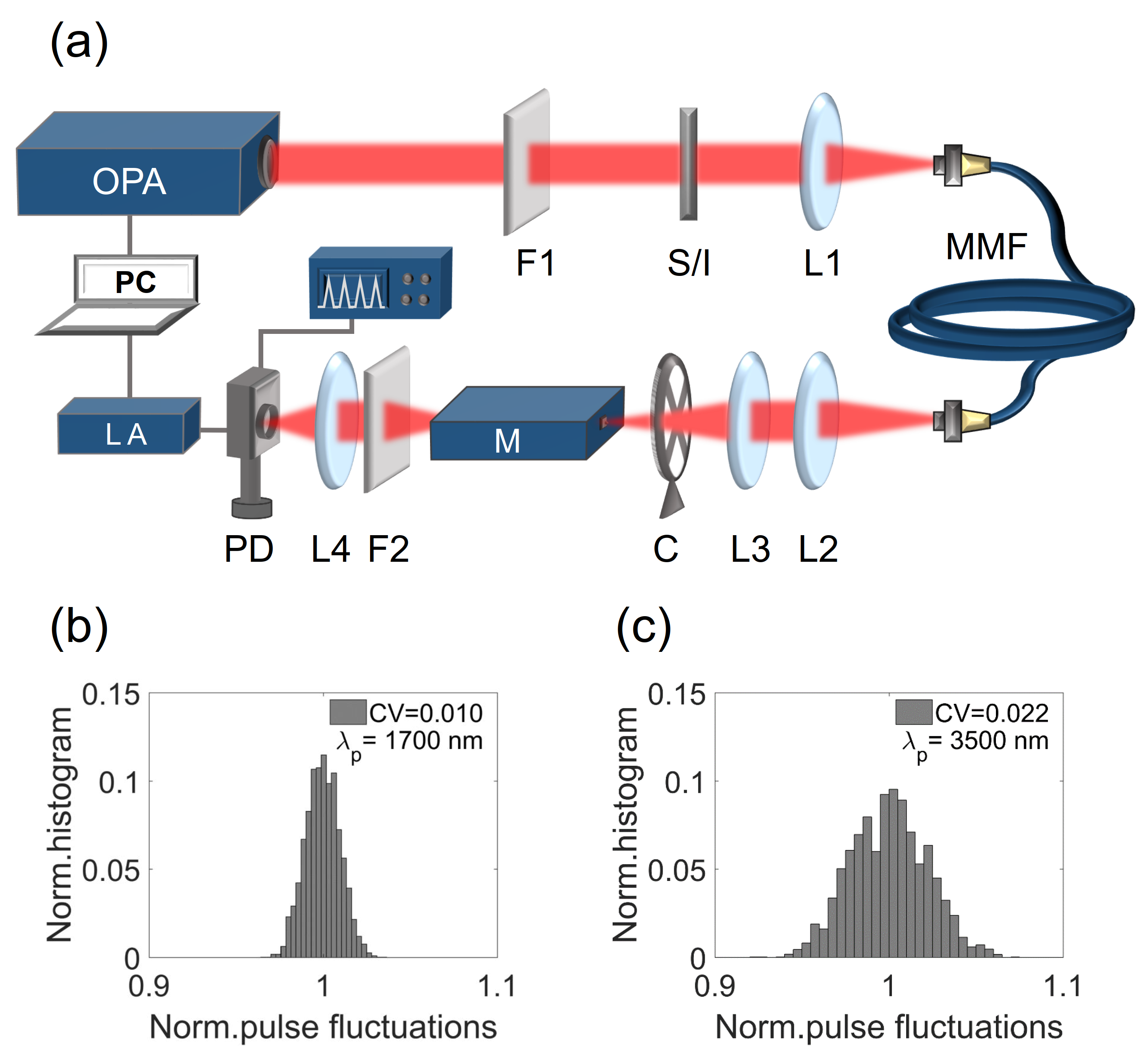}
\caption{(a) Experimental setup for mid-infrared (MIR) SC generation and intensity noise measurement. OPA: optical parametric amplifier. L1-L4: plano-convex lenses. F1-F2: long-pass filters C: chopper. M: monochromator. S/I: signal/idler isolator. LA:~ lock-in amplifier. PD: photodetector. MMF: multimode fiber. Intensity fluctuations measured over 5000 pulses corresponding to the OPA pulses at 1700 nm (b) and at 3500 nm (c).}
\label{fig:1}
\end{figure}

\begin{figure}[!b]
\centering
\includegraphics[width=\linewidth]{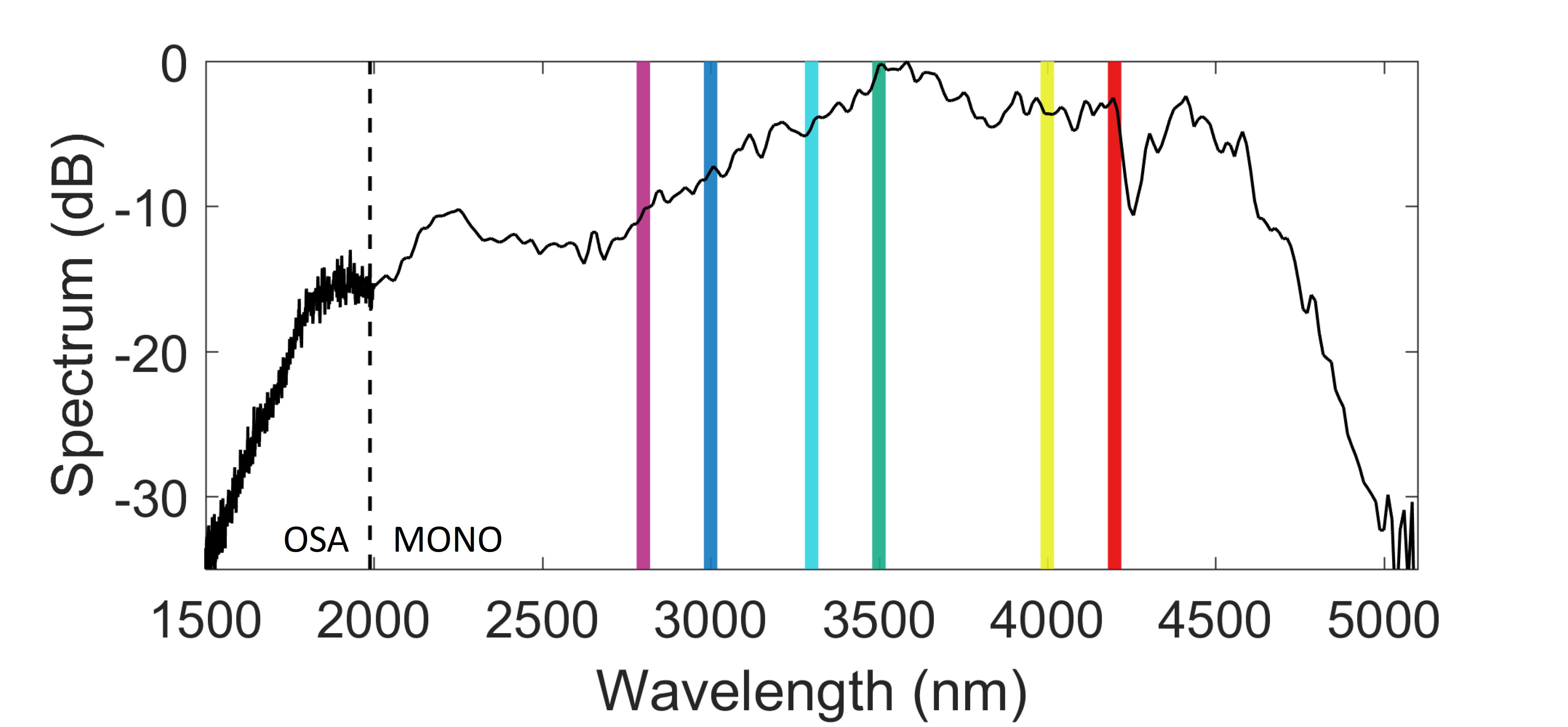}
\caption{Average SC spectrum generated in a 1-m long multimode As$_2$S$_3$ fiber with 100 $\mu$m core diameter. The dashed line marks the spectral regions measured using an OSA and monochromator. The colored areas indicate the filtered wavelength bands where the intensity fluctuations are measured.}
\label{fig:2}
\end{figure}

The experimental setup is shown in Fig.~\ref{fig:1}. Using an anti-reflection (AR) coated Si plano–convex lens with 10 cm focal length, we inject 350 fs pulses (FWHM) from an optical parametric amplifier (OPA) at a repetition of 1 MHz into a 1-m long step-index As$_2$S$_3$ multimode fiber (IRflex, IRF-S-100) with numerical aperture and diameter of NA=0.30 and 100 $\mu$m, respectively. The OPA wavelength is tuned to 3500 nm which is located in the normal dispersion regime of the fiber. The SC spectrum is measured with an optical spectrum analyzer (OSA, ANDO AQ6317B) for wavelengths in the 1000--1750 nm range and the combination of a monochromator (DK 480) and lock-in detection for wavelengths beyond 1700 nm. A maximum throughput of 60\% (calculated as a ratio of output to input power, and including coupling efficiency and attenuation along the fiber) was measured for our setup. Considering that the attenuation of the chalcogenide fiber is relatively low (about 0.05 dB/m at 2800 nm), the throughput is essentially limited by the Fresnel reflection losses due to the high refractive index contrast between the chalcogenide glass (n=2.4) and air. In order to characterize the SC pulse-to-pulse intensity fluctuations, the same monochromator was used to select wavelength-channels in narrow spectral bands of 6 nm bandwidth (experimentally measured during the calibration process of the monochromator). Light from the monochromator output was focused into a photodetector using MgF$_2$ plano–convex lens with 5 cm focal length and the electrical signal was recorded with a 9 MHz photodetector (InAsSb; PDA07P2) and 1 GHz real-time oscilloscope (LeCroy WaveRunner 6100A). Note that lock-in detection is removed for the intensity noise characterization to allow for shot-to-shot intensity measurements. 

Figure~\ref{fig:2} shows the normalized spectrum measured at the output of the 1-m long multimode As$_2$S$_3$ fiber for a pump pulse peak power of 380 kW. The SC spectrum corresponds to an average output power of 145 mW and spans from 1700 nm to 4800 nm (-30 dB bandwidth) with relatively flat spectrum (\textless10 dB variation) in the 2200--4500 nm range. The dip in the spectrum at around 4200 nm is caused by the fiber glass attenuation at this particular wavelength.

\begin{figure}[t]
\centering
\includegraphics[width=\linewidth]{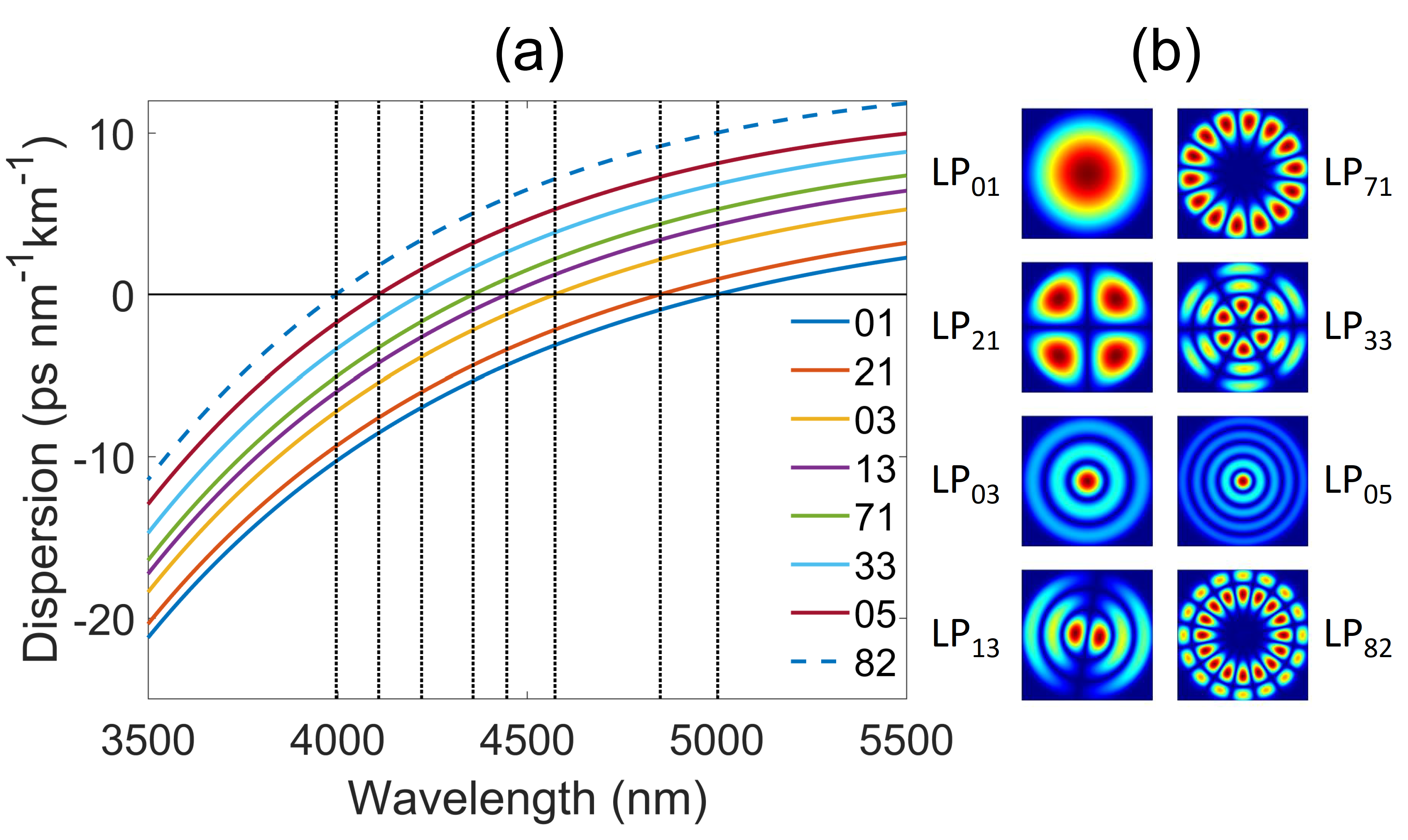}
\caption{Numerically simulated (a) dispersion and (b) mode profile of fundamental along with selected higher-order modes. Indices represent the mode number and vertical black lines mark the ZDW of the modes.}
\label{fig:3}
\end{figure}

\begin{figure}[!b]
\centering
\includegraphics[width=\linewidth]{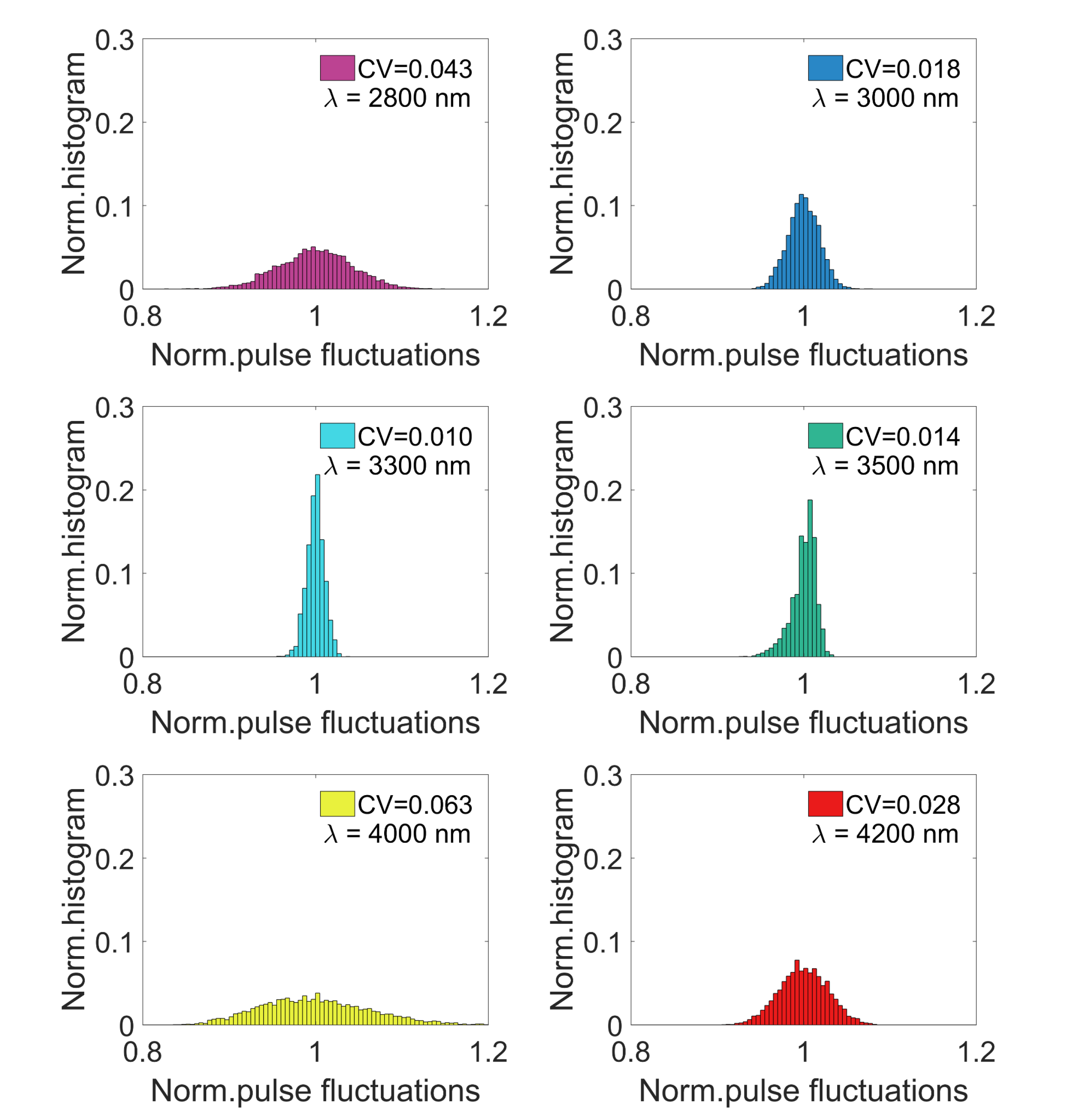}
\caption{Intensity fluctuations of the SC generated in the multimode As$_2$S$_3$ fiber measured for 5000 consecutive pulses at different wavelengths as indicated by the corresponding colors in Fig.~\ref{fig:2}. CV: coefficient of variation defined as the ratio of the standard deviation to the mean intensity.}
\label{fig:4}
\end{figure}

We performed numerical simulations of the dispersion profile associated with the fundamental and selected higher-order modes of the multimode As$_2$S$_3$ fiber. As can be seen in Fig.~\ref{fig:3}, the zero-dispersion wavelength of the fundamental mode is located at 5000 nm (close to that of the bulk glass) while that of higher-order modes decreases towards shorter wavelength with the mode order. Although we only show here selected the higher-order modes dispersion profiles, even for a higher mode order the zero-dispersion wavelength is still located above 3500 nm such that the SC is essentially generated in the normal dispersion regime with the dominant spectral broadening mechanism being self-phase modulation. This regime is less susceptible to noise amplification compared to anomalous dispersion pumping as modulation instability and bright soliton dynamics do not occur in normal dispersion and this is indeed reflected in the spectral histogram of the intensity fluctuations shown in Fig.~\ref{fig:4} measured in different wavelength bands for 5000 consecutive pulses. 

The fluctuations can be quantified by the coefficient of variation (CV) defined as the ratio of the intensity standard deviation to the mean intensity.  One can see that the fluctuations in the SC spectrum are minimal in the central part from 3000 to 3500 nm with an increase by only a factor of 2 (CV=0.043 at 2800 nm) and 3 (CV=0.063 at 4000 nm) toward the short- and long-wavelength edges, respectively, when compared to the intensity fluctuations of the pump pulses (CV=0.022, see inset in Fig.~\ref{fig:1}). Interestingly, we also note that the noise level in the vicinity of the pump wavelength is lower than that of the input pump pulses. This is because in SPM-based SC generation, the spectral intensity in the vicinity of the pump wavelength varies proportionally to the pump pulse peak power and pulse duration. In the case of a mode-locked pump laser (such as the OPA used here), the peak power and duration are anti-correlated (i.e. an increase in one causes a decrease in another and vice-versa) such that during SC generation process the noise resulting from variations in the pump pulse duration and pump pulse peak power tend to cancel each other \cite{genier2019amplitude}.

\begin{figure}[t]
\centering
\includegraphics[width=\linewidth]{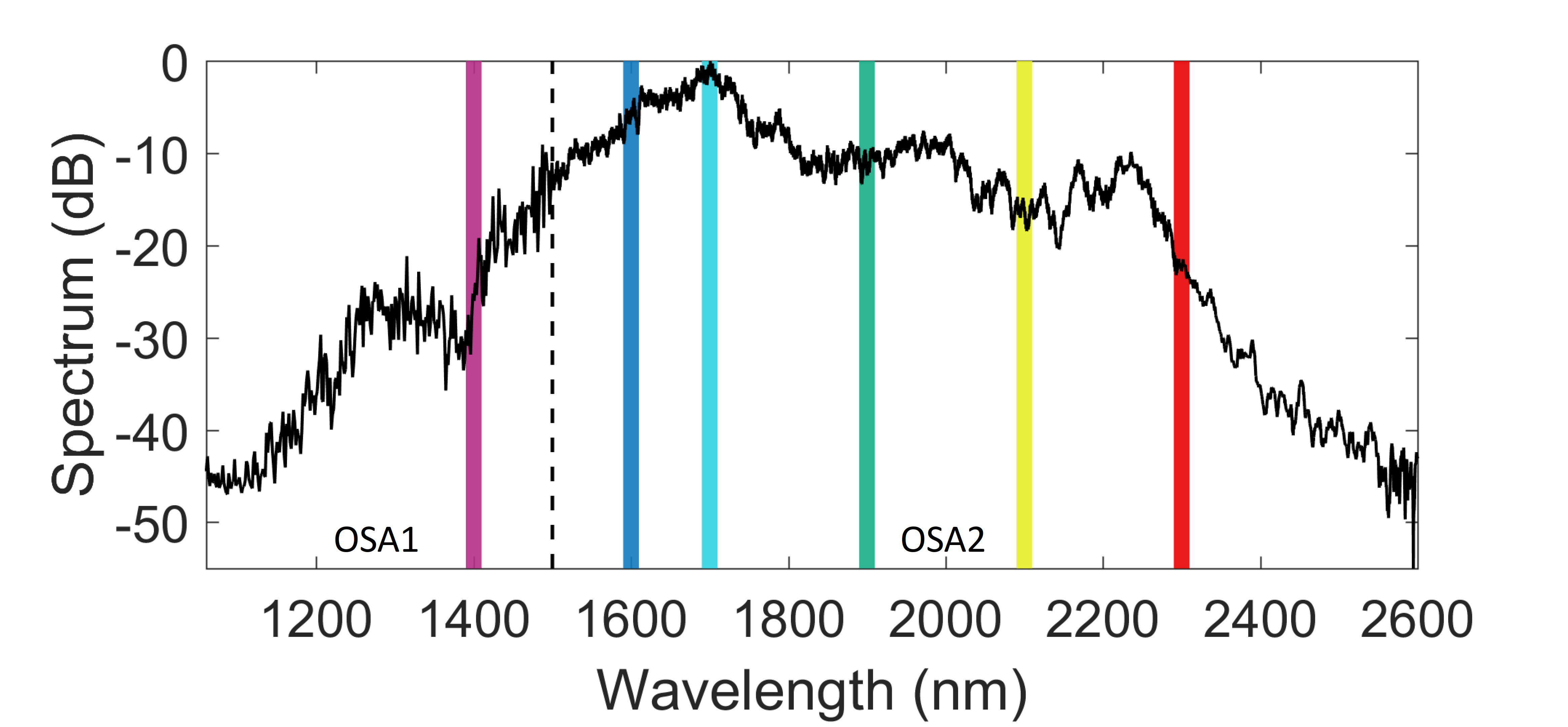}
\caption{Average SC spectrum generated in 1-m long multimode InF$_3$ fiber with 100 $\mu$m core diameter. The dashed line marks the spectral regions measured using two different OSA. The colored areas indicate the filtered wavelength bands where the intensity fluctuations are measured.}
\label{fig:5}
\end{figure}

In order to assess further the noise characteristics of the multimode SC generated in the all-normal dispersion regime against other pumping scheme, we performed a set of additional experiments in a 1-m long InF$_3$ multimode fiber using the same OPA source. The fiber has a core diameter of 100 $\mu$m similar to the As$_2$S$_3$ fiber, and a numerical aperture of NA=0.26. The OPA was tuned to 1700 nm, which is located in the normal dispersion of the fundamental mode of the InF$_3$ fiber (ZDW at around 1850 nm) but, unlike for the As$_2$S$_3$ fiber, corresponds to the anomalous dispersion regime for a large number of higher-order modes \cite{eslami2019high}. Figure~\ref{fig:5} Shows the resulting SC spectrum extending from 1200 nm to 2400 nm (-40 dB bandwidth) for an input peak power of 1.4 MW. 

\begin{figure}[!b]
\centering
\includegraphics[width=\linewidth]{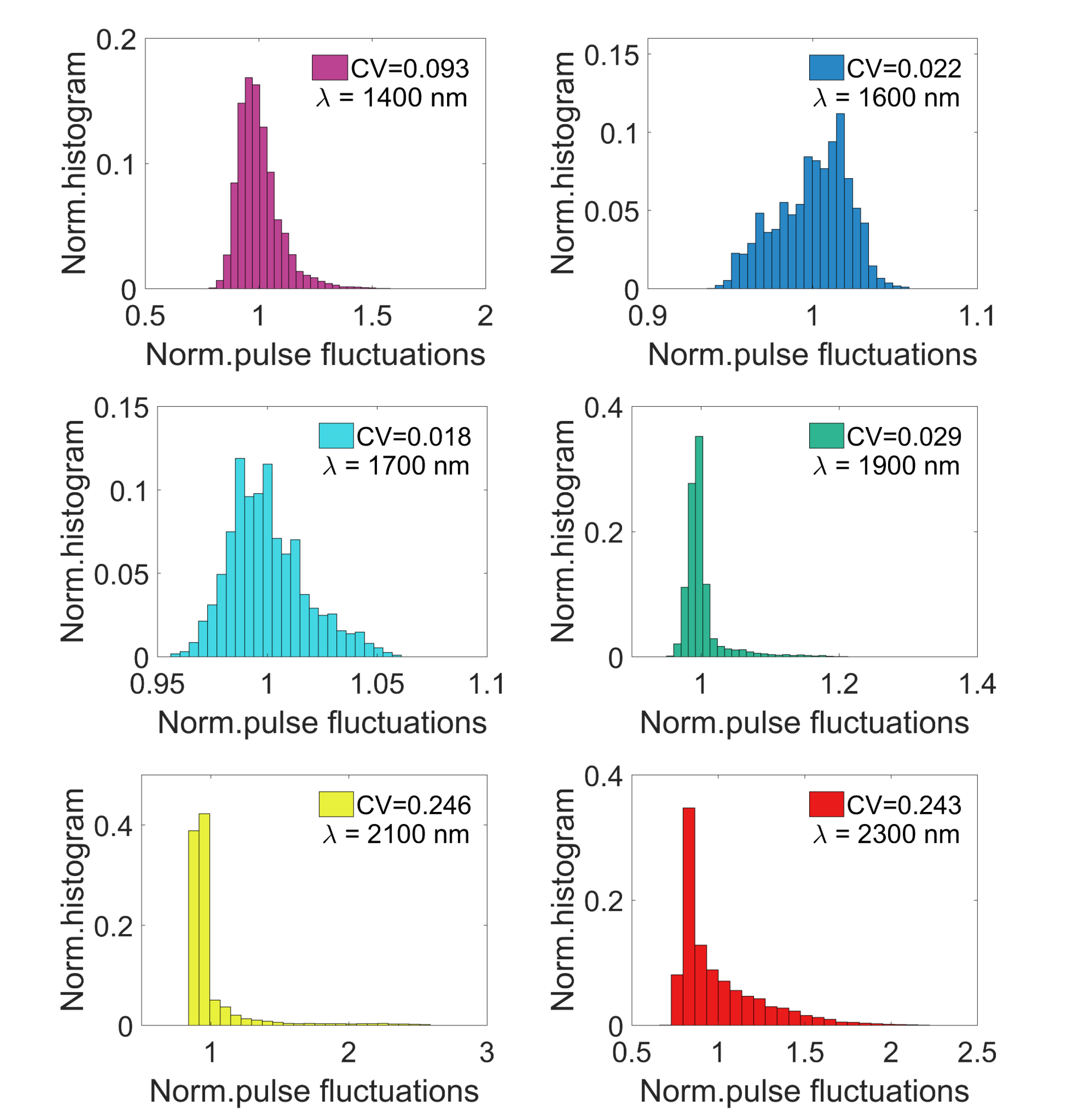}
\caption{Intensity fluctuations of the SC generated in the multimode InF$_3$ fiber measured for 5000 consecutive pulses at different wavelengths as indicated by the corresponding colors in Fig.~\ref{fig:5}. CV: coefficient of variation defined as the ratio of the standard deviation to the mean intensity.}
\label{fig:6}
\end{figure}

The intensity fluctuations of the SC were quantified in different output wavelengths using a 15 MHz photodetector (PbSe; PDA10D-EC) and the results are plotted in Fig.~\ref{fig:6}. One can see that the intensity noise in the SC spectrum is lowest in the vicinity of the pump wavelength and increases significantly toward both the short- and long-wavelength edges of the SC. The coefficient of variation is maximum at 2100 nm (CV=0.246) which is more than an order of magnitude compared to that of the pump pulses (CV=0.010, see inset in Fig.~\ref{fig:1}), which is in marked contrast with the all-normal dispersion regime SC. This can be explained from the fact that the spectral components near the pump wavelength are mostly generated by SPM but as energy is transferred to the anomalous dispersion regime of the higher-order modes the long and short-wavelength SC spectral components are generated through soliton dynamics and phase-matched dispersive waves, respectively \cite{eslami2019high}, amplifying significantly the input pulses fluctuations and leading to the observed large intensity variations.  

In conclusion, we have demonstrated the generation of a low-noise, octave-spanning mid-infrared supercontinuum in a 1-m long multimode chalcogenide fiber with 100 $\mu$m core diameter by injecting 350 fs pulses from an OPA in the normal dispersion regime of the fiber. Systematic measurements of the pulse-to-pulse fluctuations in different wavelength bands show that the noise of the input pulses is at most amplified by factor of 3. Furthermore, comparison with an octave-spanning SC generated partially in the anomalous regime of a 1-m long multimode fluoride showed that the all-normal pumping scheme using few hundred femtosecond pulses is a promising approach to generate a high-power broadband SC source for noise-sensitive applications such as remote sensing or imaging.

\bigskip

%\section*{Funding}
\noindent\textbf{{Funding.}} Horizon (2020) Framework Programme (722380).

\bigskip

 %Bibliography
%\bibliography{Noise_ref}

%\bibliographyfullrefs{Noise_ref}

%\end{document}
\bibliographystyle{ieeetr}
\bibliography{low_noise}

\end{document}